# Developing and characterizing a new-generation regolith simulant "IGCAS-AST01" for the Tianwen-2 target asteroid (469219) Kamoʻoalewa

**Pengfei Zhang[1], Zichen Wei[2], Takahiro Hiroi[3], Jin Zhao[4], Edward Cloutis[5], Guozheng Zhang[6], Marco Fenucci[7,8], Rui Li[1], Xiaojing Zhang[9], Xiaoping Zhang[6], Zhiping He[10], Yan Su[11], Yangting Lin[12], He Zhang[13], Yang Li[1]**

[1]Center for Lunar and Planetary Sciences, Institute of Geochemistry, Chinese Academy of Sciences, Guiyang, China.

[2]School of Earth Sciences, China University of Geosciences, Wuhan, China.

[3]Department of Earth, Environmental and Planetary Sciences, Brown University, Providence 02912 (RI), USA.

[4]Institute of Analysis and testing, Beijing Academy of Science and Technology, Beijing Center for Physical & Chemical Analysis, Beijing, China.

[5]Department of Geography, University of Winnipeg, Winnipeg R3B 2E9 (MB), Canada.

[6]State Key Laboratory of Lunar and Planetary Sciences, Macau University of Science and Technology, Macau, China.

[7]ESA ESRIN/PDO/NEO Coordination Centre, Largo Galileo Galilei, 1, Frascati (RM), Italy.

[8]Deimos Italia, Via Alcide De Gasperi, 24, San Pietro Mosezzo (NO), Italy.

[9]China Academy of Aerospace Science and Innovation, Beijing, China.

[10]Key Laboratory of Space Active Opto-Electronics Technology, Shanghai Institute of Technical Physics, Chinese Academy of Sciences, Shanghai, China.

[11]National Astronomical Observatories, Chinese Academy of Sciences, Beijing, China.

[12]Institute of Geology and Geophysics, Chinese Academy of Sciences, Beijing, China.

[13]China Academy of Space Technology, Beijing, China.

Corresponding author: Yang Li (liyang@mail.gyig.ac.cn)

**Key Points:**

- The regolith simulant "IGCAS-AST01" for Kamoʻoalewa is developed
- Physical, chemical, and microscopic characteristics of IGCAS-AST01 are measured and reported
- Space weathering effects on asteroid regolith properties are simulated and investigated.

**Abstract**

China plans to return samples from the near-Earth asteroid (469219) Kamoʻoalewa, which we previously identified as an LL-chondrite-compositional, highly space-weathered object with fine-grained regolith. In this study, we developed 10 mL of Kamoʻoalewa regolith simulant, designated "IGCAS-AST01", by irradiating LL5/6 chondrite (Kheneg Ljouâd) powder with a high-energy pulsed laser. We then analyzed the composition, grain size distribution, density, porosity, visible to near-infrared reflectance spectrum, thermal emission spectrum, thermal diffusivity, specific heat capacity, and microstructural features of both the fresh (unirradiated) powder and IGCAS-AST01. IGCAS-AST01 is composed of 57.8 vol.% olivine, 19.9 vol.% orthopyroxene, 5.6 vol.% diopside, 12.2 vol.% plagioclase, 2.6 vol.% troilite, and minor amounts of other phases. It has a mean size of 26.99 μm, a median size of 23.19 μm, a density of 700 kg m$^{-3}$, and a porosity of 79.1%. Additionally, IGCAS-AST01 exhibits a low reflectance of 0.1 at 0.55 μm and an extremely steep spectral slope. In the temperature range of 253.15–473.15 K, its thermal diffusivity and specific heat capacity range from 3.6–4.7 × 10$^{-6}$ m$^2$ s$^{-1}$ and 718.43–890.20 J kg$^{-1}$ K$^{-1}$, respectively. Furthermore, thick amorphous rims and abundant nanophase metallic iron particles are observed in olivine and pyroxene grains of IGCAS-AST01. These results could support the Tianwen-2 mission's payload calibration, sampling operations, on-orbit scientific data interpretation, and future sample analysis.

**Plain Language Summary**

Asteroids are small celestial bodies that record the evolutionary history of the Solar System. Analyzing returned samples provides opportunities to study the evolutionary history of asteroids in detail. To date, three international asteroid sample-return missions, specifically Hayabusa, Hayabusa2, and OSIRIS-REx, have been carried out. Now, China is preparing to carry out the fourth asteroid sample-return mission, Tianwen-2. The Tianwen-2 mission aims to return regolith samples from (469219) Kamoʻoalewa, an asteroid in Earth's orbit. To support the upcoming sampling process and data analysis, we developed a regolith simulant for Kamoʻoalewa, named "IGCAS-AST01". We further measured the physical, chemical, and microscopic characteristics of IGCAS-AST01. These results can provide key references for the implementation of the Tianwen-2 mission.

**1 Introduction**

Asteroids are small celestial bodies that orbit the Sun. As remnants of the planetary accretion process, asteroids record the early environmental information and subsequent evolutionary history of the Solar System. To date, extensive ground-based spectral surveys have revealed the compositional distribution (Gradie & Tedesco, 1982; Richard P. Binzel et al., 2004; R. P. Binzel et al., 2019; F. E. DeMeo & Carry, 2014; Marsset et al., 2022; Sanchez et al., 2024) and surface evolution mechanisms (R. P. Binzel et al., 2019; Vernazza et al., 2009; Delbo et al., 2014; Francesca E. DeMeo et al., 2023) of asteroids. However, a deeper understanding of asteroids requires the analysis of samples collected in situ. Over the past two decades, three international asteroid sample-return missions have been successfully conducted: Hayabusa to the Sq-type asteroid (25143) Itokawa, Hayabusa2 to the Cb-type asteroid (162173) Ryugu, and OSIRIS-REx to the B-type asteroid (101955) Bennu. In particular, the Hayabusa mission provided the first direct evidence linking Sq-type asteroids to ordinary chondrites (Tomoki Nakamura et al., 2011; Yurimoto et al., 2011), the Hayabusa2 mission proved the connection between Cb-type asteroids and altered CI chondrites (Yokoyama et al., 2022), and preliminary laboratory analysis suggested Bennu is also similar to CI chondrites in composition (Lauretta et al., 2024). Additionally, the combination of on-orbit observations and laboratory analyses provides key insights into the surface evolution history and mechanisms of asteroids. For example, Itokawa contains approximately 0.05 vol.% of nanophase metallic iron (npFe$^0$) (Richard P. Binzel et al., 2001; Hiroi et al., 2006), a major space weathering product in the regolith of silicate-rich bodies, and exhibits a surface exposure age ranging from $10^2$ to $10^6$ years (Bonal et al., 2015a, 2015b; Jin & Ishiguro, 2022; Keller & Berger, 2014; Koga et al., 2018; Matsumoto et al., 2018; Nagao et al., 2011). These findings suggest that Itokawa experienced moderate space weathering (R. P. Binzel et al., 2019) and non-global surface renewal driven by an impact event (Ishiguro et al., 2007; Jin & Ishiguro, 2022). For the carbonaceous asteroid Ryugu, sample analyses have revealed that it underwent a low-temperature aqueous alteration process (Hamm et al., 2022; McCain et al., 2023; T. Nakamura et al., 2022). Additionally, an upcoming asteroid sample-return mission, the Martian Moons eXploration (MMX), is scheduled to launch in 2026 and will explore the Martian moon Phobos.

Now, China has proposed a new asteroid sample-return mission, Tianwen-2. The Tianwen-2 probe is scheduled to launch in May 2025, with the goal of first returning samples from Earth's quasi-satellite (469219) Kamoʻoalewa in 2027, followed by orbiting and exploring the active main-belt asteroid 311P in 2035. The Tianwen-2 spacecraft will carry 11 payloads, including a Medium Field-of-View Color Camera, a Narrow Field-of-View Navigation Sensor, a Laser Integrated Navigation Sensor, a Multispectral Camera, a Rotational Diffraction Hyperspectral Camera, a Visible-Infrared Imaging Spectrometer (VIIS), a Thermal Radiation Spectrometer, a Detection Radar, a Magnetometer, a Charged and Neutral Particle Analyzer, and an Ejecta Analyzer (Li et al., 2024). For Kamoʻoalewa, our previous study suggested that it is a sub-hundred-meter, fast-rotating, LL-chondrite-compositional object originating from the $v_6$ secular resonance (P. F. Zhang et al., 2024). In addition, we reported that Kamoʻoalewa has developed a highly space-weathered surface composed of fine-sized regolith containing 0.29±0.05 wt.% nanophase iron (npFe$^0$) particles (P. F. Zhang et al., 2024). In this study, to support the upcoming on-orbit exploration and sampling process of the Tianwen-2 probe, we developed 10 mL of regolith simulant for Kamoʻoalewa, designated "IGCAS-AST01". We then measured its physical and chemical characteristics in detail, including composition, grain size distribution, bulk density, porosity, visible to near-infrared (VIS-NIR) reflectance spectrum, thermal radiation spectrum, thermal diffusivity, specific heat capacity, and microstructure. These results aim to provide key references for the Tianwen-2 payload calibration, sampling, on-orbit scientific data interpretation, and returned sample analysis.

**2 Methods and Results**

2.1 Simulants

In the previous study (Sharkey et al., 2021), Kamoʻoalewa was reported to exhibit a clear 1 μm absorption feature (Band I), where the Band I center matches best with that of LL ordinary chondrites among the meteorite groups we investigated (P. F. Zhang et al., 2024). Additionally, we found that a powder sample of Kheneg Ljouâd (an LL5/6 chondrite) with particle sizes <45 µm, when irradiated by a high-energy laser (40 mJ × 80 times, simulating micrometeoroid-driven space weathering), displayed a VIS-NIR spectrum matching Kamoʻoalewa (P. F. Zhang et al., 2024). Kheneg Ljouâd is a probable fresh fall (July 12, 2017) with a weathering grade of 0, so using it as the initial material for the simulant can avoid contamination from the Earth as much as possible. Based on these, we first crushed a bulk sample of Kheneg Ljouâd (Figure 1a), ground it using an agate mortar and pestle, and sieved it to obtain particles smaller than 45 µm, yielding 20 mL of light gray powder (Figure 1b). We then irradiated 10 mL of this powder with a 40 mJ × 80 times nanosecond pulsed laser at the China University of Geosciences to obtain the sample shown in Figure 1c, designated "IGCAS-AST01", which has a dark gray color. The remaining 10 mL of unirradiated powder (hereafter termed "fresh powder") was used as the reference for IGCAS-AST01.

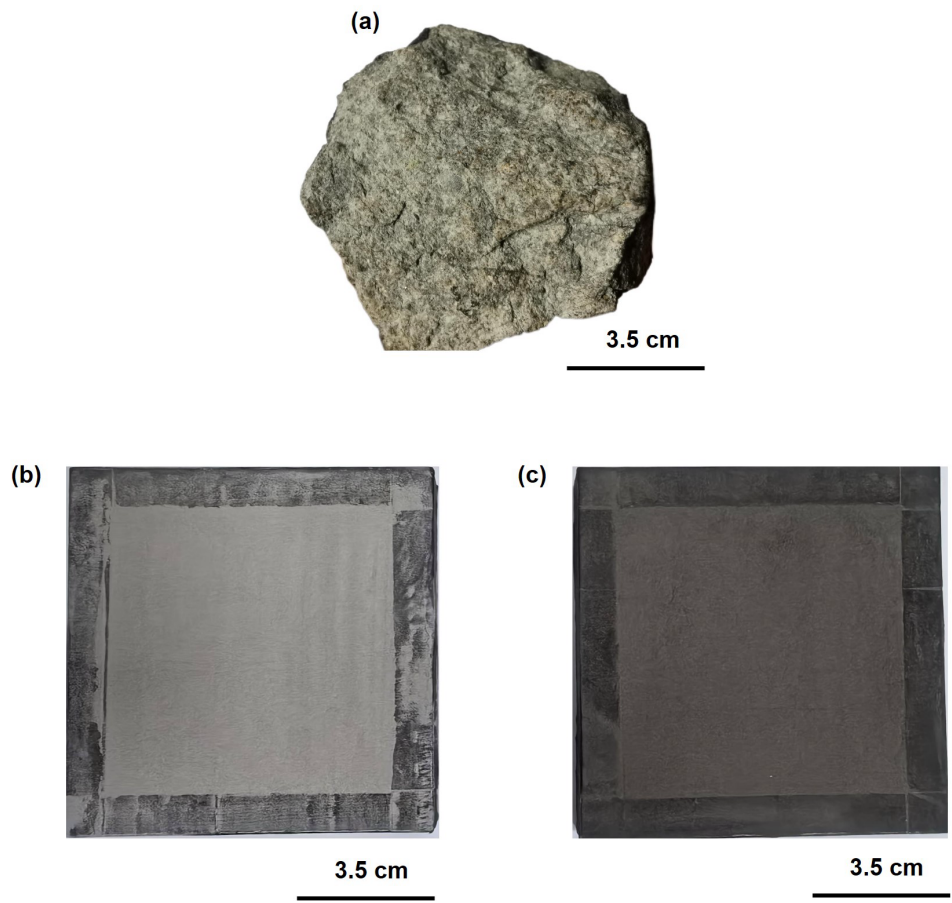

**Figure 1.** Photos of the bulk LL5/6 chondrite Kheneg Ljouâd (a), fresh powder (b), and IGCAS-AST01 (c). The fresh powder exhibits a light gray color, while IGCAS-AST01 appears a dark gray color.

2.2 Composition

We first measured the composition of the bulk sample from Kheneg Ljouâd (Figure 1a) to estimate the composition of IGCAS-AST01. The bulk sample represents the unweathered state of Kamoʻoalewa and was chosen because it is more convenient to measure. The main difference between the bulk sample and IGCAS-AST01 is that the latter contains npFe$^0$ particles and agglutinates; however, in general, their mineral compositions are otherwise similar. We used the TESCAN Integrated Mineral Analyzer (TIMA) at Nanjing Hongchuang Geological Exploration Technology Service Co., Ltd. to characterize the mineral types and contents. It is equipped with a MIRA-3 scanning electron microscope and four X-ray energy dispersive spectroscopy detectors to obtain elemental information.

From our analysis, the Kheneg Ljouâd consists of 57.73 vol.% olivine, 19.87 vol.% orthopyroxene, 5.61 vol.% diopside, 12.23 vol.% plagioclase, 2.62 vol.% troilite, 0.05 vol.% nickel-bearing troilite, 0.56 vol.% tetrataenite, 0.76 vol.% chromite, 0.03 vol.% ilmenite, 0.12 vol.% orthoclase, and 0.38 vol.% apatite. Although we did not analyze its geochemical composition, the measurement results from the Meteoritical Bulletin Website (https://www.lpi.usra.edu/meteor/) can serve as supplementary data: olivine (Fa$_{31.0\pm0.2}$), low-Ca pyroxene (Fs$_{25.0\pm0.4}$Wo$_{2.1\pm0.2}$), high-Ca pyroxene (Fs$_{10.7}$Wo$_{43.1}$ and Fs$_{11.0}$Wo$_{43.0}$), Feldspar (Ab$_{84.4\pm2.2}$An$_{10.6\pm0.3}$Or$_{5.0\pm2.3}$), and tetrataenite (Fe$_{42.9\pm0.2}$Co$_{2.1\pm0.1}$Ni$_{54.9\pm0.2}$).

2.3 Grain Size Distribution

We used a laser particle size analyzer at the Institute of Analysis and Testing, Beijing Academy of Science and Technology to characterize the grain size distribution of the fresh powder and IGCAS-AST01. The results showed that the fresh powder had a mean size of 29.17 µm and a median size of 26.65 µm (Figure 2a), whereas IGCAS-AST01 showed a mean size of 26.99 µm and a median of 23.19 µm (Figure 2b). Additionally, after laser irradiation, the proportion of coarser particles decreased (Figures 2a–b). These findings are consistent with the concept that space weathering driven by micrometeoroid impacts could lead to particle size reduction (Pieters & Noble, 2016). Notably, Figures 2a–b show a small fraction of particles with sizes exceeding 45 µm, likely due to some particles having a long axis length >45 µm but pass through the screen in the direction of the short axis (<45 µm).

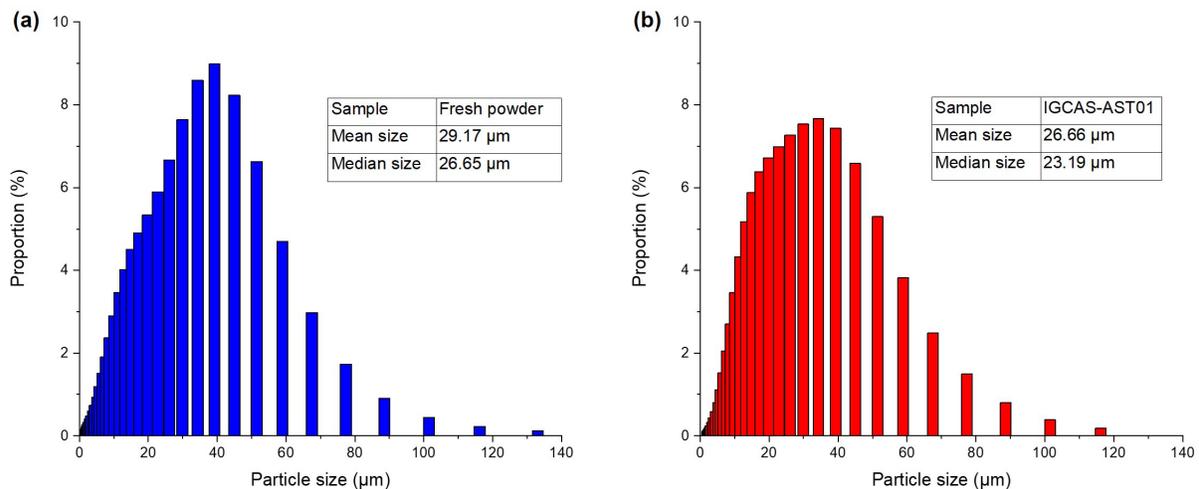

**Figure 2.** Particle size distribution of fresh powder (a) and IGCAS-AST01 (b). IGCAS-AST01 exhibits a slightly smaller mean size and median size compared to fresh powder.

2.4 Bulk Density and Porosity

We used a precision balance to measure the mass of a standard-sized cuboid (10 mm × 10 mm × 3 mm) of the Kheneg Ljouâd block and determined its bulk density to be 3357 kg m$^{-3}$. We also placed loosely packed fresh powder and IGCAS-AST01 into a transparent sapphire sample cell (10 mm × 10 mm × 3 mm) and measured their masses. From these measurements, we obtained their bulk densities of 760 kg m$^{-3}$ and 700 kg m$^{-3}$, respectively. Assuming zero porosity for the bulk sample, the porosities of the fresh powder and IGCAS-AST01 were calculated to be 77.4% and 79.1%, respectively. The porosity exhibited a marginal increase after irradiation, likely due to the formation of agglutinates and the growth of micropores induced by laser irradiation.

2.5 VIS-NIR Reflectance Spectra

We used the UV-VIS-NIR bidirectional spectrometer and Thermo Nexus 870 FT-IR spectrometer at Reflectance Experiment Laboratory (RELAB) in Brown University to measure the reflectance spectra of fresh powder and IGCAS-AST01. The former measures the spectrum from 0.3 to 2.6 µm, while the latter covers 1 to 99.7 µm. We then digitally merged the spectra from the two wavelength ranges together. Since the VIIS onboard the Tianwen-2 spacecraft only covers the wavelength range of 0.45–4.5 µm, in order to facilitate comparison with data obtained in the future, we only show the spectra of fresh powder and IGCAS-AST01 in the same wavelength range here. During measurements of two wavelength ranges, a Spectralon standard (SRS-99 by Labsphere) and a brushed diffuse gold board were used as references, respectively. All measurements were conducted in an atmospheric environment with an incidence angle of 30°, an emission angle of 0°, and a phase angle of 30°.

As shown in Figure 3a, the fresh powder exhibits a high visible reflectance (0.282 at 0.55 µm) and a slightly red VIS-NIR spectral slope, while IGCAS-AST01 has a low visible reflectance (0.103 at 0.55 µm) and an extremely red spectral slope. Additionally, the fresh powder shows two wide and deep absorption bands near 1 µm (Band I) and 2 µm (Band II), which are characteristic of olivine and pyroxene (Burns, 1970; Clark, 1957). We calculated the centers and depth of Band I and Band II by removing the continuum (note that to be consistent with previous studies, the Band II continuum's right endpoint is set at 2.45 µm) and performing third- and fourth-order polynomial fittings over the bottom third of each band. Thus, centers of Band I and Band II of fresh powder are 0.996±0.001 µm (with a depth of 28.3%) and $1.952^{+0.004}_{-0.002}$ µm (with a depth of 6.6%), respectively. After irradiation, Band I and Band II become weaker, with centers at 0.988±0.002 µm (with a depth of 15.9%) and $1.951^{+0.005}_{-0.004}$ µm (with a depth of 4.2%), respectively. We also calculated the area ratio of Band II to Band I (BAR) using a self-developed MATLAB program "BA calculation program", and the BARs of fresh powder and IGCAS-AST01 are 0.25 and 0.41, respectively. Additionally, both the fresh and irradiated spectra exhibit a wide absorption near 3 µm, indicating adsorbed atmospheric water. The fresh powder and IGCAS-AST01 also display two narrow absorptions at 3.43 µm and 3.52 µm, which are attributed to carbonates (Pearson & Calvin, 2014), likely formed due to weathering on Earth. However, these two absorptions are weak and nearly disappear in IGCAS-AST01. As shown in Figure 3b, the IGCAS-AST01 exhibits a good match with the ground-based observed spectrum of Kamo'oalewa, indicating that IGCAS-AST01 can simulate the surface state of Kamo'oalewa well.

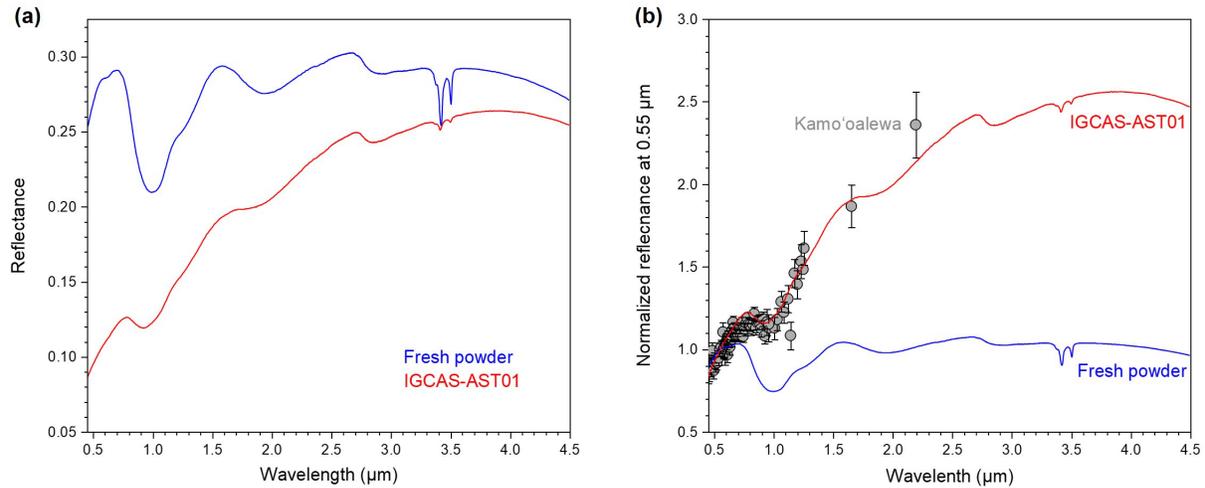

**Figure 3.** Spectral comparison among fresh powder, IGCAS-AST01, and Kamo'oalewa. (a) VIS-NIR reflectance spectra of fresh powder and IGCAS-AST01, where IGCAS-AST01 exhibits the lower reflectance, weaker absorption, and a redder spectral slope than fresh powder. (b) Normalized reflectance spectra of fresh powde, IGCAS-AST01, and Kamo'oalewa, where IGCAS-AST01 matches Kamo'oalewa well. The spectral data for Kamo'oalewa are cited from Sharkey et al. (2021).

2.6 Thermal Emission Spectra

We converted reflectance to emissivity using the Kirchhoff law (Emissivity = 1 – Reflectance). To facilitate comparison with future data obtained by the Thermal Radiation Spectrometer onboard Tianwen-2 spacecraft (which has a wavelength range of 5–50 μm), we only show the spectra in the 5–50 μm wavelength range here.

As shown in Figure 4, fresh powder and IGCAS-AST01 exhibit similar curve shapes and positions for $CF_1$ (first Christiansen feature peak), $CF_2$ (second Christiansen feature peak), TF (Transparency feature peak), and $RF_2$ (second Reststrahlen feature peak). Fresh powder exhibits slightly lower emissivity than IGCAS-AST01. For fresh powder, $CF_1$ is located at 1153 cm$^{-1}$ (8.67 μm), $CF_2$ at 652 cm$^{-1}$ (15.34 μm), TF at 768 cm$^{-1}$ (13.03 μm), $RF_1$ (first Reststrahlen feature peak) at 931 cm$^{-1}$ (10.74 μm), and $RF_2$ at 519 cm$^{-1}$ (19.28 μm). For IGCAS-AST01, $CF_1$ is located at 1153–1155 cm$^{-1}$ (8.66–8.67 μm), $CF_2$ at 652 cm$^{-1}$ (15.34 μm), TF at 773 cm$^{-1}$ (12.93 μm), $RF_1$ at 887 cm$^{-1}$ (11.27 μm), and $RF_2$ at 519 cm$^{-1}$ (19.28 μm). After irradiation, $CF_1$, $CF_2$, and $RF_2$ remained unchanged. However, laser irradiation altered the curve shape near $RF_1$ and caused TF to shift slightly toward shorter wavelengths.

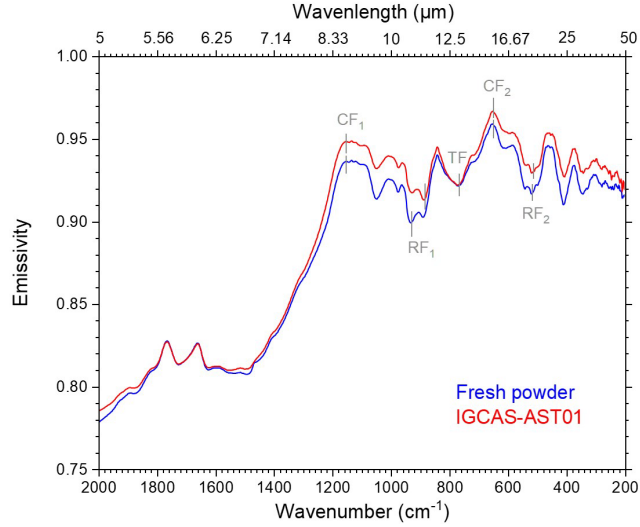

**Figure 4.** Thermal emission spectra of fresh powder and IGCAS-AST01. Fresh powder exhibits slightly lower Emissivity than IGCAS-AST01.

2.7 Thermal Diffusivity, Specific Heat Capacity, Thermal Conductivity, and Thermal Inertia

In the temperature ($T$) range of 253.15–473.15 K, we measured the thermal diffusivity of a 10 mm × 10 mm × 3 mm block sample, fresh powder, and IGCAS-AST01 in a sapphire sample cell using the flash method. The principle involves irradiating a homogeneous sample of a specific size with a high-intensity, short-duration pulsed laser at a set temperature. The lower surface of the sample absorbs energy, leading to an instantaneous temperature increase, and as the hot end, it transfers energy to the cold end (upper surface) via approximately ideal one-dimensional heat conduction. The temperature rise curve at the center of the sample's upper surface is recorded, and the half-rise time is obtained by fitting this curve. Subsequently, the thermal diffusivity ($\alpha$) was calculated from the half-rise time. Meanwhile, differential scanning calorimetry (DSC) was used to measure the isobaric specific heat capacity ($C_P$) of these samples. Thus, combined with the measured bulk density ($\rho$), thermal conductivity ($K$) was obtained using equation (1):

$$K = \rho \alpha C_P \tag{1}$$

Additionally, thermal inertia ($I$) was calculated via equation (2):

$$I = \sqrt{\rho K C_P} \tag{2}$$

As shown in Table 1, the thermal diffusivity of the block sample is one order of magnitude higher than that of fresh powder and IGCAS-AST01, with IGCAS-AST01 exhibiting a slightly higher value than fresh powder. The block sample shows the lowest specific heat capacity, followed by IGCAS-AST01, while fresh powder exhibits the highest value. The thermal conductivity of the block sample is one order of magnitude higher than that of fresh powder and IGCAS-AST01, with IGCAS-AST01 being slightly lower than fresh powder. The thermal inertia of the block sample (1418.56 to 1527.00 J m$^{-2}$ K$^{-1}$ s$^{-1/2}$) is one order of magnitude higher than that of fresh powder and IGCAS-AST01, with IGCAS-AST01 again slightly lower than fresh powder. These results suggest that space weathering may slightly reduce the heat transfer capacity of asteroid regolith.

**Table 1.** Bulk density ($\rho$), thermal diffusivity ($\alpha$), specific heat capacity ($C_P$), thermal conductivity ($K$), and thermal inertia ($I$) of the block LL5/6 chondrite Kheneg Ljouâd, fresh powder, and IGCAS-AST01.

| Samples | $\rho$ (kg m$^{-3}$) | T (K) | $\alpha$ (m$^2$ s$^{-1}$) × 10$^{-7}$ | $C_P$ (J kg$^{-1}$ K$^{-1}$) | K (W m$^{-1}$ K$^{-1}$) | I (J m$^{-2}$ K$^{-1}$ s$^{-1/2}$) |
|---|---|---|---|---|---|---|
| Block sample | 3357 | 253.15 | 4.99 | 598.20 | 1.0021 | 1418.56 |
| | | 273.15 | 4.72 | 638.08 | 1.0110 | 1471.63 |
| | | 293.15 | 4.23 | 692.01 | 0.9827 | 1510.89 |
| | | 313.15 | 4.03 | 725.32 | 0.9813 | 1545.73 |
| | | 353.15 | 3.70 | 757.63 | 0.9410 | 1547.07 |
| | | 373.15 | 3.52 | 771.18 | 0.9113 | 1535.95 |
| | | 423.15 | 3.20 | 799.28 | 0.8586 | 1517.84 |
| | | 473.15 | 3.08 | 819.62 | 0.8475 | 1527.00 |
| Fresh powder | 760 | 253.15 | 0.35 | 710.46 | 0.0199 | 101.02 |
| | | 273.15 | 0.28 | 755.47 | 0.0161 | 96.07 |
| | | 293.15 | 0.38 | 807.31 | 0.0233 | 119.60 |
| | | 313.15 | 0.33 | 848.44 | 0.0213 | 117.14 |
| | | 353.15 | 0.44 | 905.52 | 0.0303 | 144.36 |
| | | 373.15 | 0.34 | 928.29 | 0.0240 | 130.09 |
| | | 423.15 | 0.34 | 977.79 | 0.0253 | 137.02 |
| | | 473.15 | 0.42 | 992.58 | 0.0317 | 154.60 |
| IGCAS-AST01 | 700 | 253.15 | 0.36 | 718.43 | 0.0181 | 95.52 |
| | | 273.15 | 0.35 | 752.61 | 0.0184 | 98.56 |
| | | 293.15 | 0.37 | 785.41 | 0.0203 | 105.75 |
| | | 313.15 | 0.39 | 804.88 | 0.0220 | 111.27 |
| | | 353.15 | 0.04 | 851.36 | 0.0238 | 119.19 |
| | | 373.15 | 0.46 | 878.87 | 0.0283 | 131.95 |
| | | 423.15 | 0.44 | 895.32 | 0.0276 | 131.46 |
| | | 473.15 | 0.47 | 890.20 | 0.0293 | 135.09 |

2.8 Microscopic Features

We also used the FEI Scios Dual Beam scanning electron microscope (SEM) at the Institute of Geochemistry, Chinese Academy of Sciences to observe the morphological characteristics of grains of fresh powder and IGCAS-AST01. Additionally, we investigated the simulated space weathering features in olivine and pyroxene grains of IGCAS-AST01 using the FEI Talos F200X field-emission transmission electron microscope (TEM) at the Sinoma Institute of Materials Research (Guangzhou) Co., Ltd.

As a result, fresh powder grains exhibit sharp edges (Figure 5a). After laser irradiation, the grains become rounder, and numerous agglutinates appear (Figure 5b). We also observed amorphous rims at the edges of olivine and pyroxene grains in IGCAS-AST01, where numerous npFe$^0$ particles are embedded in the lower and upper parts of the rims (Figures 5c–d). Additionally, the amorphous rim of olivine grains is thicker (115−225 nm) than that of pyroxene (71−182 nm) (Figures 5c–d). Agglutinates, amorphous rims, and npFe$^0$ particles, as typical space weathering products, have been widely found in Chang'e-5 lunar soil (Gu et al., 2022) and returned particles from Itokawa (Takaaki Noguchi et al., 2014). The comprehensive reproduction of these microstructural features in IGCAS-AST01 demonstrates its superior fidelity in simulating space weathering effects compared to previous simulants.

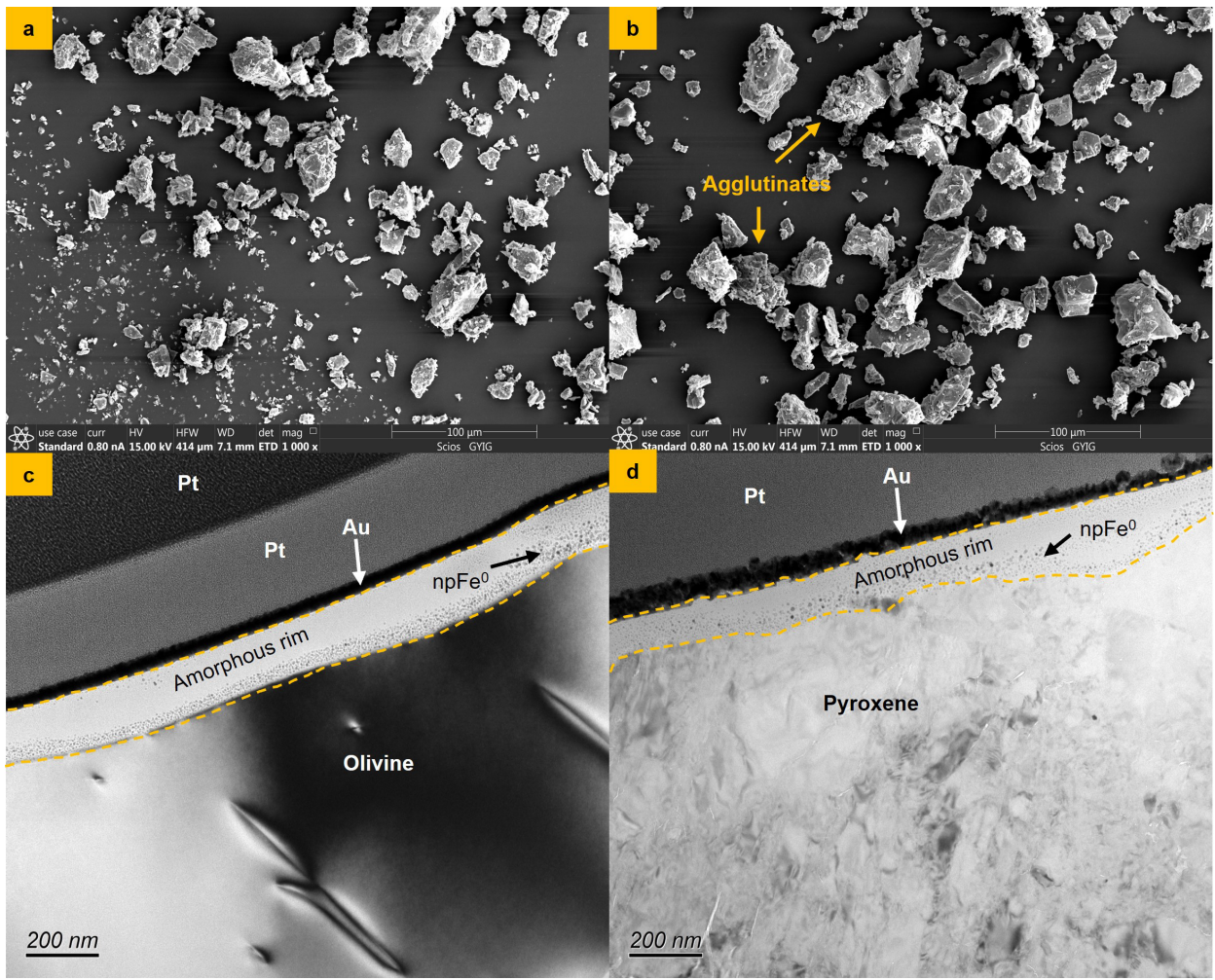

**Figure 5.** Microscopic features of fresh powder and IGCAS-AST01. (a–b) SEM secondary electron images of fresh powder (a) and IGCAS-AST01 (b). Fresh powder grains show sharp edges, while IGCAS-AST01 has a large number of agglutinates. (c–d) TEM bright field images of focused ion

beam (FIB) sections sampled from olivine (c) and pyroxene (b) in IGCAS-AST01. Amorphous rims and abundant npFe$^0$ particles were found in olivine and pyroxene of IGCAS-AST01.

## 3 Discussion and conclusions

### 3.1 Implications for Tianwen-2 mission

The development and characterization of simulants facilitate (1) payload calibration in the laboratory, (2) testing of sampling protocols, and (3) prediction of returned sample characteristics. Unlike the Hayabusa, Hayabusa2, and OSIRIS-REx missions, Tianwen-2 will be the first mission to conduct sampling and return operations on a sub-100-meter, fast-rotating asteroid. The potential low surface gravity and rapid rotation of Kamoʻoalewa pose significant challenges for sample collection. Although a previous simulant for Kamoʻoalewa was produced by mixing terrestrial minerals (X. Zhang et al., 2021), it is still necessary to develop a simulant that more closely resembles the actual state of Kamoʻoalewa's regolith. Accordingly, our IGCAS-AST01 uses a corresponding meteorite (specifically, LL chondrite) as the initial material, with controlled particle size and laser irradiation processing, distinguishing it from conventional terrestrial mineral mixtures. Our IGCAS-AST01 shows a good match with the ground-based observed spectrum of Kamoʻoalewa. The thermal inertia of IGCAS-AST01 (95.52–135.09 J m$^{-2}$ K$^{-1}$ s$^{-1/2}$) is also close to the newly observed result of Kamoʻoalewa ($150^{+90}_{-45}$ to $181^{+95}_{-60}$ J m$^{-2}$ K$^{-1}$ s$^{-1/2}$) reported by Fenucci et al., (2025). These indicate that IGCAS-AST01 can simulate the regolith state of Kamoʻoalewa well.

The Tianwen-2 spacecraft is designed with three sampling modes: gas-assisted rolling sweeping, scooping, and gripping. Currently, the grain size of Kamoʻoalewa's regolith is estimated to be fine, ranging from 100 μm to 3 mm (Fenucci et al., 2025) or smaller than 1–2 cm (Ren et al., 2024; P. F. Zhang et al., 2024). This means that the first two sampling modes may be more suitable for collecting IGCAS-AST01-like fine particles, while the gripping mode is suitable for collecting centimeter-sized debris. Collecting both fine particles and coarse debris will enable comparative laboratory analyses to investigate the space weathering history of Kamoʻoalewa.

The Tianwen-2 probe will hover 20 km above the surface of Kamoʻoalewa and collect the first VIS-NIR reflectance spectrum using the VIIS. A comparison between the collected spectrum of Kamoʻoalewa and our measured spectra of fresh powder and IGCAS-AST01 will be beneficial for quickly detecting Kamoʻoalewa's composition, space weathering degree, and spectral type. In addition, the thermal data we measured for the block sample, fresh powder, and IGCAS-AST01 can serve as reference input parameters for calculating Kamoʻoalewa's surface thermal inertia and estimating the regolith thickness.

### 3.2 Effects of space weathering on asteroid regolith properties

As a common process that occurs on the surface of celestial bodies, space weathering has been widely studied for its effects on material microstructure and VIS-NIR spectra (Lantz et al., 2017; Matsuoka et al., 2020; Pieters & Noble, 2016; P. Zhang et al., 2022). As observed in the fresh powder and IGCAS-AST01 (Figures 3–5), for asteroids with ordinary chondrite compositions, space weathering produces amorphous rims and npFe$^0$ particles, reduces the reflectance, diminishes the band depth, and increases the spectral slope (Hiroi et al., 2006; T. Noguchi et al., 2011; Pieters & Noble, 2016; P. Zhang et al., 2022). However, the effects of space weathering on other physical parameters have rarely been investigated.

Our particle size distribution analyses of the fresh powder and IGCAS-AST01 in Figure 2 indicate that laser irradiation slightly reduces the particle size, consistent with the view that micrometeoroid impacts could break grains into smaller-sized particles or dust (Pieters & Noble, 2016). The results in Section 2.4 indicate that space weathering of silicate-rich asteroidal regolith may decrease bulk density and increase porosity. Since the depth of amorphous rims (tens to hundreds of nanometers) is much smaller than mid-infrared wavelengths (several to tens of micrometers), previous views tend to be that space weathering would not significantly alter mid-infrared spectra. However, our measurements of the fresh powder and IGCAS-AST01 in Figure 4 indicate that space weathering may slightly shift the TF center toward shorter wavelengths and the $RF_1$ center toward longer wavelengths in silicate-rich asteroids. Previous series of laser irradiation experiments on olivine gave different results, namely that the TF center remained unchanged, slightly shift toward the shorter or longer wavelength (Yang et al., 2017). The mechanisms underlying these shifts on IGCAS-AST01 are currently unclear, perhaps related to mineral amorphization and (or) the formation of $npFe^0$ particles. Additionally, as revealed in Table 1, laser irradiation slightly increases the thermal diffusivity and decreases the specific heat capacity. Laser irradiation also reduces the powder density, further resulting in a slight decrease in thermal conductivity and thermal inertia. These subtle variations suggest that using parameters from unweathered samples may lead to the overestimation of the thermal conductivity and thermal inertia of silicate-rich asteroids.

## Acknowledgments

We would like to thank Yazhou Yang and Yusi Luo for beneficial discussions. This work was supported by the National Natural Science Foundation of China (42441804, 42273042, U24A2008, and 42303041); Youth Innovation Promotion Association CAS awards; "From 0 to 1" Original Exploration Cultivation Project, Institute of Geochemistry, Chinese Academy of Sciences (DHSZZ2023.3); Bureau of Frontier Sciences and Basic Research, CAS (QYJ-2025-0103); Guizhou Provincial Foundation for Excellent Scholars Program (GCC [2023] 088); Provincial Key Research and Development Plan Projects of Heilongjiang (2024ZXDXB52); Innovation and Development Fund of Science and Technology of Institute of Geochemistry, Chinese Academy of Sciences; Guizhou Province Basic Research Program Project (QKHJC-ZK (2023)-General 473); Natural Sciences and Engineering Research Council of Canada (RGPIN-2021-02995) and Canadian Space Agency (22EXPOSIWI). RELAB is a multiuser facility supported by NASA Planetary Science Enabling Facilities program 80NSSC23K0198.

## Conflict of Interest

The authors declare no conflicts of interest relevant to this study.

## Data Availability Statement

The VIS-NIR and thermal emission spectral data, as well as the MATLAB program to calculate the band area in this study are available at the Zenodo database (https://doi.org/10.5281/zenodo.15174637).